\documentclass[sigconf]{acmart}
\usepackage{enumitem}
\usepackage{amsmath}
\usepackage{lipsum}

\AtBeginDocument{%
  }

\copyrightyear{2026}
\acmYear{2026}
\setcopyright{cc}
\setcctype{by}
\acmConference[ITiCSE 2026]{Proceedings of the 31st ACM Conference on Innovation and Technology in Computer Science Education V.1}{July 13--15, 2026}{Madrid, Spain}
\acmBooktitle{Proceedings of the 31st ACM Conference on Innovation and Technology in Computer Science Education V.1 (ITiCSE 2026), July 13--15, 2026, Madrid, Spain}
\acmPrice{}
\acmDOI{10.1145/XXXXXXX.XXXXXXX}

\begin{document}

\title[Self-Regulated Personal Contracts as a Harm Reduction Approach \\ to Generative AI in Undergraduate Programming Education]{Self-Regulated Personal Contracts as a Harm Reduction Approach to Generative AI in Undergraduate Programming Education}


\author{Aadarsh Padiyath}
\affiliation{%
  \institution{University of Michigan}
  \city{Ann Arbor}
  \country{USA}}
\email{aadarsh@umich.edu}

\author{Jessica Shen}
\affiliation{%
  \institution{University of Michigan}
  \city{Ann Arbor}
  \country{USA}}
\email{jesishen@umich.edu}

\author{Barbara Ericson}
\affiliation{%
  \institution{University of Michigan}
  \city{Ann Arbor}
  \country{USA}}
\email{barbarer@umich.edu}


\begin{abstract}

Students learning programming exercise agency in deciding when and how to use GenAI tools like ChatGPT. However, this agency is often implicit and shaped by deadline pressure and peer behavior rather than explicit and conscious learning goals. We designed a GenAI Contract grounded in harm reduction and self-regulated learning theory to scaffold intentional decision-making: students articulated personal learning goals, created usage guidelines, and reflected on alignment at strategic points across an eleven-week semester. The contract was non-binding and graded only for completion, emphasizing self-awareness over enforcement. We implemented this with N=217 students in an intermediate Python course. For students still forming their relationship with GenAI, it worked, as 58\% of students reported the intervention changing their thinking and created helpful accountability structures. However, awareness didn't always translate to sustained behavior change. Some students who valued their guidelines still abandoned them under various pressures. Maintaining guidelines required constant self-control across hundreds of decisions, while using GenAI freely requires none. Many students could not sustain this burden despite this self-awareness. We discuss supporting student agency when GenAI tools and learning goals create tension.
\end{abstract}



\maketitle

\section{Introduction}

Students learning programming today face a stream of decisions \cite{gandhi2025s}: Should I ask ChatGPT to explain this error message? Should I use Copilot's suggestion, or write the code myself? Should I generate a solution to check my work, or would that short-circuit my learning? These decisions happen constantly, often shaped by factors beyond the tools themselves \cite{bernstein2025beyond}: peer behavior and career anxieties \cite{padiyath2024insights}, late-night exhaustion and deadline pressure \cite{denny2024desirable}, as well as genuine curiosity \cite{boguslawski2025programming}. Most of these choices are implicit; students rarely pause to ask whether their Generative AI (GenAI) use aligns with their learning goals \cite{prather2024widening, padiyath2024insights}.

Instructors face their own choices. Some prohibit GenAI entirely \cite{lau2023ban}, others embrace it enthusiastically, believing `resistance is futile' \cite{vadaparty2024cs1, lau2023ban}. Some chart a middle ground, placing restrictions on certain uses \cite{prather2023robots, ali2025analysis}. These are not neutral responses. Each policy choice influences student behavior and classroom culture \cite{ali2025analysis}. Yet any policy should acknowledge the reality that programming students are not a monolith. Some are training for software development careers, others are curious about programming but do not expect themselves to code professionally \cite{cunningham2022bringing}. Some arrive confident in their own programming skills; others are struggling with basic computational thinking. One-size-fits-all rules cannot serve this diverse population equally well \cite{padiyath2024have}. Different students may need different GenAI approaches based on their unique goals, motivations, and futures. 
Yet GenAI use is shadowed by judgment: students worry GenAI reliance will be seen as incompetence \cite{padiyath2024insights, yilmaz2023augmented}, while instructors debate which uses constitute academic dishonesty \cite{lau2023ban, prather2023robots}. This climate of judgment makes it difficult to have honest conversations about what students are actually doing and why.

\textit{Harm Reduction} \cite{marlatt1996harm} offers a philosophical framework designed for exactly this tension: when something is widely accessible, when people may use it regardless of rules, and when judgment may drive behavior underground rather than preventing it. Originating in public health, the philosophy acknowledges that people make choices regardless of official policy \cite{marlatt1996harm}. Rather than mandating abstinence or morally condemning use, harm reduction focuses on supporting better decision-making. For example, comprehensive sex education provides information and access to contraception rather than relying solely on abstinence-only education \cite{naisteter2010minimizing, kohler2008abstinence}. Applied to our context, this means recognizing that students encounter GenAI tools as readily available options \cite{ko2025student, hou2024effects}, that 'appropriate use' varies by student and goal \cite{padiyath2024insights}, and our role as educators is helping them navigate their choices thoughtfully rather than policing behavior.

However, thoughtful navigation requires substantial self regulatory skill \cite{gandhi2025s}. Students need to monitor whether a strategy serves their goals, recognize when it is leading them astray, and adjust their approach accordingly \cite{gandhi2025s, wang2025chatgpt}. This kind of intentional practice takes time and experience to develop, and cannot be reduced to a universal rule ("always use GenAI for X, never for Y"). 
What we can do is create opportunities for students to build this capacity towards aligning their GenAI use (and non-use) with their personal goals rather than following external rules they may resent or ignore.

This paper describes our implementation of a \textit{"GenAI Contract"}: a structured reflection tool grounded in self-regulated learning theory and harm reduction principles. Students set their own learning goals, create personal usage guidelines, and reflect multiple times throughout the semester on whether their choices align with their aspirations. Importantly, the contract is non-binding and not enforced. We designed four reflection prompts across eleven weeks, grading for completion rather than content.

We implemented this contract reflection process with $N$=217 students in an intermediate Python course. Our goals were (1) to understand whether students would meaningfully engage with structured reflection about their GenAI choices; (2) to identify patterns in how they set goals and adjusted their approach over time; and (3) to determine what aspects of this intervention actually supported or hindered the development of intentional practice. This experience report shares what we learned: what worked, what didn't, and how other instructors might adapt this approach.

\section{Prior and Related Work}

\subsection{Designing for Agency: Self-Regulation and Harm Reduction}

Problem-solving and critical thinking skills are essential to learn in programming, but many studies show a tension between building these skills and using AI for learning assistance \cite{bernstein2025beyond, kirkpatrick2025challenging}. The recent adoption of GenAI tools has made it increasingly easy for students to bypass active problem-solving steps. Although these tools can be helpful, they can also lead to over-reliance or over-trust in AI's help \cite{bernstein2025beyond}. AI being easily accessible makes it difficult for students to judge when to rely on the tool versus their own reasoning, often leading to a lack of intention when using the tools \cite{prather2024widening, padiyath2024insights}. This has led to a widening gap between students who can and can't manage these tools effectively \cite{prather2024widening}. We draw from robust theoretical frameworks to guide students toward intentional AI use and increase self-awareness in learning practices.

We draw on Self-Regulated Learning (SRL): the active process of executing cognitive control during a task \cite{loksa2022metacognition}. SRL depends on two foundational elements: metacognition and self-efficacy. Metacognition is a student's knowledge about their own cognitive control and effective strategies. This is important in the context of GenAI as students may over-trust the tool, leading to relying on the tool rather than building their own understanding. Self-efficacy is the belief in one's capacity to successfully execute a specific task \cite{loksa2022metacognition, margulieux2024self, bandura1982self}. This belief is crucial because self-efficacy influences a student's persistence and internal motivation when faced with challenges \cite{loksa2022metacognition}. Without the students' awareness of their use of GenAI and the confidence in their own abilities, students lack the necessary tools to confidently navigate the learning process. To operationalize SRL, we use \citeauthor{zimmerman2002becoming}'s continuous three-phase model, which breaks the learning process down into forethought (planning), performance (monitoring), and self-reflection (judgement) \cite{zimmerman2002becoming}. By implementing this loop, SRL can put students in charge of how they think and use GenAI, preventing them from falling into a passive role when using the tool. This approach directly scaffolds the self-awareness and metacognition required to intentionally use GenAI tools in a responsible manner in their learning process.

Our approach is also rooted in harm reduction philosophy (HR) \cite{marlatt1996harm}. We chose this approach because GenAI tools are readily accessible, and research shows that students may turn to them because of time pressure and low motivation \cite{ko2025rethinking, hou2024effects}, regardless of policies that try and control student behavior. HR acknowledges that GenAI use can be both helpful and harmful for learning and therefore shifts the focus. Rather than promoting abstinence or morally condemning use, it focuses on encouraging more responsible and intentional use, which respects student agency and supports informed decision-making. This intervention aims to close the gap between students who can and can't manage the tool effectively.

Finally, implementation intention plans combine the SRL framework and HR into an actionable practice \cite{gollwitzer1999implementation}. These plans are action-planning strategies that link situational cues to a desired goal \cite{gollwitzer1999implementation}. For example, a student's plan may be "If I am stuck on an error for more than 30 minutes, then I will ask GenAI to explain the concept and an example without asking for the exact solution". This commitment statement allows the student to maintain intentionality and discourages turning to GenAI too early. By having students formalize their planned GenAI usage into "contracts" ahead of time, we operationalize what Padiyath \cite{padiyath2024have} theorized as necessary: making these implicit goals explicit. This allows students to consciously align their GenAI use with their learning goals and directly builds off the SRL forethought phase. Furthermore, revisiting and revising the contract scaffolds the self-monitoring phase and allows students to make adjustments to their strategies as they learn and gain experience. Lastly, asking students to evaluate their usage aligns with the self-reflection phase. By integrating our theoretical model into a structured tool in the form of a GenAI contract, we investigate how to empower programming students to become more self-aware of their learning process.

\subsection{Programming Students and GenAI Tools}

How programming students actually use generative AI is more nuanced than the utopian or dystopian narratives suggest \cite{padiyath2024insights}. Many students do report genuine benefits from GenAI tools. Many use them to quickly debug errors, understand error messages, or explore alternative approaches to a problem \cite{padiyath2024insights, prather2023robots}. For some, the tools lower barriers to larger projects as well as expressing themselves through programming \cite{vadaparty2024cs1, zastudil2023generative}. However, recent research also reveals patterns of use that may be troubling \cite{padiyath2024insights, prather2024widening, bernstein2025beyond}. Students who use GenAI early and frequently report decreased confidence in their own abilities \cite{padiyath2024insights, margulieux2024self}. Some describe a passivity that develops as they copy GenAI-created solutions without engaging with the underlying logic or code. Notably, many students either do not recognize this \cite{prather2024widening}, or only recognize it after it affects their academic performance \cite{padiyath2024insights}. 

This pattern suggests the tools' affordances for speed and productivity may work against learning and understanding. After all, GenAI tools were not specifically designed with educational contexts in mind. Their affordances (generating complete, expert-like code instantly) combined with formal institutional structures and curricula, create a powerful pull toward productivity-focused use rather than learning-focused use \cite{padiyath2024insights}. While feeling efficient, productivity-focused use may undermine students' learning goals. Students with substantial self-regulatory capacity can use GenAI in ways aligned with learning, but doing so requires conscious resistance to the tool's default affordances \cite{gandhi2025s, prather2024widening}. Yet general self-regulation skills may not transfer automatically, as \citeauthor{margulieux2024self} found no relationship between students' self-regulation strategies and their AI use \cite{margulieux2024self}. In essence, students need to develop intentional use (and deliberate non-use) of GenAI: the ability to consciously choose how GenAI fits into their learning goals, rather than defaulting to passive use patterns \cite{padiyath2024have}.

\subsection{Approaches to Supporting Intentional Use}

Educators and researchers have proposed various integrations of GenAI in programming education to support intentional use and learning, yet most focus on modifying the tool itself -- restricting access, scaffolding its outputs, or teaching students how to use it more effectively \cite{kazemitabaar2024codeaid, liffiton2023codehelp, liu2024teaching}. These tool-focused approaches often rest on the assumption that if we design better tools or teach better tool use, students will learn better. However, technology adoption and use are shaped by both technical \textit{and} human factors including student goals, peer influence, institutional contexts, and crucially, student agency and intentionality \cite{padiyath2024have, padiyath2024insights}.

SRL research offers a more learner-centered lens, but applying it in programming education can be cognitively demanding. \citeauthor{parham2010empirical} found that programming students tend to skip important metacognitive steps like planning and monitoring in favor of immediately writing code \cite{parham2010empirical}. \citeauthor{wang2025chatgpt}'s controlled study integrating ChatGPT with \citeauthor{zimmerman2002becoming}'s SRL framework found improvements in students' motivation and self-efficacy, but no significant gains in learning outcomes, which the authors attribute to cognitive overload from managing SRL strategies alongside programming itself \cite{wang2025chatgpt}. Critically, when scaffolding was minimal, students bypassed self-regulatory prompts to complete tasks more quickly \cite{wang2025chatgpt}. These findings suggest that even when SRL is embedded within tool design, GenAI's frictionless affordances make reflective prompts easy to bypass.
This matters because modifying tools alone cannot account for how students actually decide to use GenAI \cite{kazemitabaar2024codeaid}, and whether that use actually aligns with their learning goals \cite{prather2024widening, padiyath2024insights}. Few interventions directly address and support the metacognitive and self-regulatory dimensions of this decision-making \cite{bernstein2025beyond}.
Drawing on self-regulated learning theory and harm reduction frameworks, our approach places self-regulation outside the tool interaction entirely, making reflection the explicit focus rather than an add-on to an already demanding task.

\section{Intervention Design and Context}

\subsection{Course Context}

Our intervention took place in Fall 2025 in an intermediate-level Python programming course at a large public research university in the Midwest United States. The course enrolled N=217 students taught by one instructor alongside twelve undergraduate TAs. This course focused on the following Python skills: functions, conditionals, object-oriented programming basics, debugging, Git version control, Regular Expressions, I/O, APIs, and databases. Eight homework assignments and two projects gave students opportunities to practice these skills. Two proctored midterm exams assessed students' abilities without external resources including GenAI. To take this course, students completed at least one of several prerequisite introductory programming courses.

Our student population contained approximately 60\% Information Science majors (for whom the course was required), 10\% Computer Science majors, and 30\% from various other programs. The Information Science students typically envisioned futures as UX designers, data scientists, or "conversational programmers" \cite{cunningham2022bringing}. Computer Science students, whose required sequence used C++, often take this course specifically to learn Python. Due to these varying degrees of expertise, a one-size-fits-all policy on GenAI use would fit no one particularly well.
Of note, our syllabus stated: "You can also use generative AI like ChatGPT, but must list that as a thing you worked with and what it helped you with."

\subsection{GenAI Contract Design}

Given this context (diverse student goals, permissive GenAI policy, and free access to available AI tools), we needed an intervention that could support students in making thoughtful choices when using GenAI tools. 

Four design principles, each grounded in our theoretical frameworks, shaped our contract. First, it would be non-binding, explicitly presented as a reflection tool rather than an enforced contract, reflecting HR's `come as you are' philosophy of supported decision-making over judgment \cite{marlatt1996harm}. Second, it would be goal-driven: students articulate personal learning objectives before considering GenAI strategies, operationalizing \citeauthor{zimmerman2002becoming}'s forethought phase \cite{zimmerman2002becoming}; the if-then guideline format then draws on implementation intention theory \cite{gollwitzer1999implementation} to translate those goals into concrete, situationally-anchored action plans. Third, it would support iterative reflection, with students revisiting and revising guidelines at strategic points across the semester, scaffolding \citeauthor{zimmerman2002becoming}'s performance monitoring and self-reflection phases \cite{zimmerman2002becoming} rather than producing a one-time artifact. Fourth, it would afford a diversity of approaches, consistent with both HR's recognition that appropriate use varies by individual \cite{marlatt1996harm} and with the reality that our students had genuinely different learning goals.

The contract structure translated these into four reflection prompts across eleven weeks (see Figures \ref{fig:contract12} and \ref{fig:contract34}). In Week 2, students completed Steps 1 and 2 together as a single Google Doc assignment. Steps 3 and 4 were distributed as text for students to copy into their existing contract document. Step 3 was distributed in Week 5, after students had taken their first midterm. This timing was chosen due to its finding as a important time for reflection on GenAI usage \cite{padiyath2024insights}. The final reflection, Step 4, came in Week 11 after the second midterm. By this point, students had completed much coursework and taken both midterms. Each reflection was worth a small number of completion points to incentivize participation (1.5\% of their final grade). We graded only for completion, never for content.

\begin{figure*}[t]
\centering
\small
\begin{minipage}{0.95\textwidth}
\fbox{\begin{minipage}{0.98\textwidth}

\begin{center}
\textbf{\large GenAI Contract}
\end{center}

\vspace{0.5em}

\noindent\textbf{Background}

\noindent Students don’t just “use” GenAI tools like ChatGPT – they decide when, how, and why to use them. These choices are often implicit, shaped by goals, habits, and pressures. This contract is a simple reflection tool to help you make those decisions more deliberate.

\textbf{Important}: 
\textcolor{red}{\textbf{This contract will not be enforced.}}
It is only for your personal reflection and growth. The goal isn’t to get it “right” the first time; it is to increase your awareness of how GenAI fits with your learning.

\vspace{0.5em}

\noindent\textbf{Step 1: Your Goals in This Course}

\noindent Take a minute and actually read through the course description for this course, taken from the syllabus.

\vspace{0.5em}


\hspace{5em}\begin{minipage}{0.7\textwidth}\textbf{COURSE DESCRIPTION}

This course will focus on giving you a strong background in data-oriented programming, in the Python programming language. It is intended for students who have completed an introductory programming course and are moving on to the next step in a data-oriented fashion. Learning objectives for this course include:

\begin{itemize}
    \item Develop intermediate programming skills in Python

    \item Practice using basic data structures (lists, tuples, dictionaries)

    \item Learn how to use a terminal window to run a Python program locally on your computer

    \item Familiarity with the basics of object-oriented programming: objects, classes, and inheritance

    \item Develop debugging and testing skills

    \item Use a distributed code repository (git and GitHub).

    \item Develop experience with using pattern matching (RegEx)

    \item Work with data from a variety of sources (files, web, APIs, JSON, and databases)

    \item Work with relational databases with SQL
\end{itemize}
\end{minipage}

\vspace{0.5em}

\noindent Before thinking about how you might use GenAI, pause to consider what you want to get out of this course. Your goals might relate to your major, career plans, GPA, personal interests, skills you want to build, etc.

Be specific. Vague statements like “I want to do well in this class” aren’t useful as something you can act on and reflect on later.

Start with at least one clear personal goal below, and feel free to list more if you like. You can use this format to describe your goals if you want: In this course, I <\textit{your goal here}> because <\textit{why that goal makes sense for you}>.

\vspace{0.25em}

\noindent\textbf{Step 2: Your Personal Approach to GenAI}

\noindent Think about how you might use GenAI – or not – throughout this course. The goal is to \textbf{make your thinking explicit} so you can revisit it later. \textbf{You can describe situations where you plan to use GenAI, situations where you plan not to, or a mix.}

\textbf{These guidelines are for you and you alone.} You do not have to follow these guidelines perfectly; they exist only to help you reflect and make intentional choices.
Focus on your reasoning – what matters to you, and why – rather than trying to follow a “right” pattern.

Start with \textbf{at least one clear guideline about your personal approach to GenAI}. You can add as many additional guidelines as you wish. Just make sure each one is relevant to you and your goals in this course. You can use this format to describe your guidelines if you wish: I will <\textit{the way you will use/not use GenAI}> when <\textit{condition/constraint in which you will/won’t use GenAI}> because <\textit{reasoning that makes sense for you}>.

\end{minipage}}
\end{minipage}
\vspace{-1mm}
\caption{The GenAI Contract template distributed to students through a Google Doc during Week 2.}
\label{fig:contract12}
\vspace{-4mm}
\end{figure*}

\begin{figure*}[t]
\centering
\small
\begin{minipage}{0.95\textwidth}
\fbox{\begin{minipage}{0.98\textwidth}

\noindent\textbf{Step 3: Post-Midterm 1 Reflection}

\begin{enumerate}[leftmargin=*, nosep, itemsep=2pt]
    \item Describe how you have actually used GenAI in this class so far (during homework, studying, debugging, learning concepts, etc.):
    \begin{itemize}
        \item What types of situations prompted you to use it?
        \item What generally did you ask it to do or help you with?
    \end{itemize}
    
    \item Looking back at the goal(s) you created in Step 1, do you think your GenAI usage helped or hurt you in working toward those goals? Has your goal(s) changed since the beginning of this course? Explain your reasoning.

    \item What changes, if any, do you want to make to your GenAI usage guideline(s) you created in Step 2 going forward?
    
    You can use this format to describe your updated guidelines if you wish: I will <\textit{the way you will use/not use GenAI}> when <\textit{condition/constraint in which you will/won’t use GenAI}> because <\textit{reasoning that makes sense for you}>. OR I'm keeping my guidelines the same because <\textit{explain how your current approach is working for your goals}>.

\end{enumerate}

\vspace{0.25em}

\noindent\textbf{Step 4: Post-Midterm 2 Reflection}

\begin{enumerate}[leftmargin=*, nosep, itemsep=2pt]
    \item Before you reflect on the semester thus far, think about your recent GenAI usage:
    \begin{itemize}
        \item In between Midterm 1 and Midterm 2, how often did you use GenAI for this class? [Not at all - Very frequently each week (10+ sessions)]

        \item How does this compare to your usage earlier in the semester prior to Midterm 1? [Much less than before Midterm 1 - Much more than before Midterm 1]
    \end{itemize}
    
    \item You've now taken both midterms:
    \begin{itemize}[nosep, itemsep=1pt]
        \item How confident do you feel in your programming skills compared to the start of the semester? [much less confident - much more confident]
        \item Describe times when you chose to use GenAI, and separate times when you chose other resources (peers, lectures, office hours, etc.). Explain the reasoning behind your choices. What role do you think your GenAI usage played compared to other resources?
        \item How do you feel about how you managed your GenAI use this semester? (Proud / Satisfied / Conflicted / Regretful / Other) Explain why.
    \end{itemize}
    
    \item Look back at the guidelines you created in Step 2 and/or 3.
    \begin{itemize}[nosep, itemsep=1pt]
        \item Did you stick to them, modify them informally, or abandon them?
        \item Were your guidelines effective strategies for your own learning goals you created in Step 1? Why or why not?
    \end{itemize}
    \item You might not have this contract reflection structure in future classes, but you'll keep making choices about GenAI.
    \begin{itemize}[nosep, itemsep=1pt]
        \item Based on what you learned about yourself this semester, what's ONE insight about yourself as a learner that will shape how you approach GenAI tools going forward?
        \item Complete this planning statement: "When learning programming in the future, I will use GenAI when \underline{\hspace{2cm}} because \underline{\hspace{2cm}}. I will NOT use GenAI when \underline{\hspace{2cm}} because \underline{\hspace{2cm}}."
    \end{itemize}
    \item This contract served as a self-regulation tool, as it asked you to plan (Steps 1-2), monitor (Step 3), and reflect (this step).
    \begin{itemize}[nosep, itemsep=1pt]
        \item Did having this contract process change your GenAI usage or thinking about it? If so, how? If not, why not?
        \item If you were re-designing this tool/process to help future students self-regulate their GenAI use when learning programming, what would you change?
    \end{itemize}
\end{enumerate}

\end{minipage}}
\end{minipage}
\vspace{-1mm}
\caption{The GenAI Contract template distributed to students across two time points: Step 3 during Week 5, post-Midterm 1; and Step 4 during Week 11, post-Midterm 2. Students received Steps 3 and 4 as a text to copy/paste into their existing GenAI Contract Google Doc containing Steps 1 and 2.}
\label{fig:contract34}

\vspace{-4mm}
\end{figure*}

\subsection{Data Collection and Analysis}

This methodology was approved by our institution's review board. Our primary data source for analysis was student contract submissions. To understand what worked and what didn't, the research team read through all contract submissions, taking detailed memo notes on patterns across students, surprising responses, contradictions and tensions, confusions, and interesting quotes. Our analysis was guided by our three goals: Did students engage meaningfully with the reflection prompts? What patterns emerged in how students aligned (or did not align) their GenAI use with their learning goals? And what aspects of the contract design seemed to support or hinder the development of intentional practice? As this is an experience report rather than a formal qualitative research study, we did not develop a formal codebook nor calculate inter-rater reliability. Instead, we report our observations on what worked and what didn't as instructor-researchers who read all students contracts. Statistics were generated by the first author.

\section{Results and Takeaways}

Student engagement with the contract process was relatively high across all three steps. Of the 217 enrolled students, 211 completed Steps 1-2 (97\%), 200 completed Step 3 (92\%), and 185 completed Step 4 (85\%). While the quality of reflections varied, 
most responses contained sufficient detail for us to understand students' experiences and GenAI practices. The following subsections describe patterns in students' responses. Since a single student could describe multiple experiences (e.g., finding the contract valuable while also abandoning guidelines under pressure) the percentages below are not mutually exclusive. It is also worth considering our analysis makes use of self-reported data, which has its own limitations.

\subsection{What Worked: Helping Novices Form an Approach}

More than half the students (58\%, N=108/185) reported that the contract changed how they thought about GenAI Usage. Many of these students described arriving to the course without existing approaches to using AI tools. They had not given much thought to when, why, or how they engaged with GenAI in their learning process -- they were simply using it as available. One student captured this shift: \textit{"Having this contract process definitely changed how I think about using GenAI. It made me more aware of when and why I was using it, and helped me realize that setting personal boundaries with AI tools can actually make me a stronger learner."} 

These students created concrete if-then rules tied to specific situations: time-based triggers (\textit{"I will use GenAI after trying for 5 minutes"}, \textit{"I will ask GenAI ... if I am unable to solve the problem by myself within 30-40 minutes"}), context-based conditions (\textit{"when office hours aren't available"}, \textit{"when I am working alone"}), and action-based constraints (\textit{"I will use AI when I need a concept to be explained in very simple terms"}, \textit{"I will use GenAI, but carefully word the prompt so that GenAI only helps me without giving me the answer"}). 
The written guidelines functioned as accountability structures. One student wrote, \textit{"Having a contract changed my brain when considering using AI, I’d always remember the rules I set for myself."} 

In contrast, some students (25\%, N=46/185) arrived with established approaches to GenAI use. One student wrote, \textit{"I have always used it as a learning tool rather than a `cheating' tool."} Another noted: \textit{"I already knew what boundaries I wanted to set for myself for AI use."}. Other students had made deliberate choices to use GenAI more extensively. One explained: \textit{"I will use GenAI to find a solution ... because in the workforce, I will just be using AI anyway and correcting the work of the computer not the other way around so why not learn to use it."} Another simply stated: \textit{"[I use GenAI] since I am only given a short amount of time. I rather get the points and not understand than not get the points and understand. ... I just want to pass the class and get my degree to get a good paying job."}
For these students -- whether they planned restricted or extensive use, whether they framed their approach through learning goals or pragmatic GPA/career goals -- the contract documented existing practices rather than creating new ones. Thus, when asked if the contract changed their thinking, they reported "No."

This pattern aligns with implementation intention research. \citeauthor{gollwitzer2006implementation}'s work demonstrates that if-then plans are most effective for initiating new behaviors, but less effective when strong habits (whether productive or unproductive) already exist \cite{gollwitzer2006implementation}. 
Implementation intentions are traditionally tools for behavior initiation, not habit modification. Students without established practices used the contract to operationalize \citeauthor{zimmerman2002becoming}'s SRL forethought phase \cite{zimmerman2002becoming} -- translating their personal learning goals to concrete action plans. The iterative checkpoints created opportunities for performance monitoring and self-reflection. But students who had already systematized their GenAI use found the contract confirmatory rather than useful. As one put it: \textit{"I didn't really reflect or use this contract that much. I am generally pretty good about my Ai usage and I don’t think I needed this contract."} 

From a harm reduction perspective, this outcome supports the approach. Harm reduction emphasizes "meeting people where they are" and supporting decision-making without imposing external control \cite{marlatt1996harm}. Students with established practices had already developed the decision-making capacity the intervention aimed to support, and students without those practices gained scaffolding they needed.

\textbf{Takeaway:} Self-regulation contracts work as helpful scaffolding for students initiating new behaviors, not for modifying existing habits. Instructors might see differential impact based on students prior experiences with intentional GenAI use.

\subsection{What Didn't Work: Forgetting about the Contract and Time Pressure}

A smaller group of students (13\%, N=23/185) reported forgetting about the contract between checkpoints. Students created guidelines in Week 2, returned to them in Week 5 after Midterm 1, and again in Week 11 after Midterm 2. In the weeks between, the students were not required to read or recall their guidelines. Students wrote: \textit{"I honestly haven’t thought about this contract much."}, and \textit{"It's very out of sight out of mind, once I finish it it's done and I never think about it again."}. Some treated it as perfunctory: \textit{"honestly, this contract just seemed like another assignment to me. There wasn’t ever really any impact."} The contract lived in a Google Doc visited three times across the course, not where students made actual decisions about GenAI use. Students themselves identified this design flaw, suggesting \textit{"weekly check-ins"} and ways to be \textit{"constantly reminded about"} their guidelines.

More significantly, some students who valued their guidelines reported abandoning them under various pressures (11\%, N=21/185). One wrote: \textit{"I use Gen AI when I am starting working on an assignment way too last minute, and feel pressured to just submit it on time."} Another captured the tension: \textit{"There were absolutely situations where GenAI saved me a stupid amount of time and effort, like with debugging. At the same time, I do feel there were some instances where the "time-saving" argument was used in place of just critically thinking about what needs to be done."}

\textbf{Takeaway:} Awareness and willpower alone may not be sufficient for sustained self-regulation under pressure. Future implementations could: (1) integrate guidelines into workflow where students make decisions, (2) increase reflection frequency to catch misalignment earlier, (3) more explicitly scaffold the revision process, and (4) consider assignment pacing/structures that reduce crunch time.

\subsection{Student Suggestions for Redesigning the Contract}
\label{subsection:suggestions}

When asked for feedback on the contract process, students suggestions clustered around three main ideas: increasing frequency and visibility, providing better scaffolding, and building in technical supports.

\textbf{More frequent touchpoints.} As discussed previously, students proposed bi-weekly or even weekly check-ins instead of midterm-based reflections: \textit{"I would include short weekly check-ins or reflection prompts instead of one big reflection"} and \textit{"I would hold students even more accountable by maybe adding a reflection every week to talk about their use of AI weekly."} Students recognized that infrequent reflections allowed problematic patterns to continue unchecked. Although this may lead to intervention fatigue, more frequent touchpoints could address the forgetting problem by keeping guidelines active rather than allowing them to disappear for weeks at a time.

\textbf{Concrete examples and peer learning.} Students wanted more examples about what responsible use looked like. \textit{"I'd include more examples or short reflections from past students about how they used GenAI successfully,"} one wrote. Another suggested, \textit{"I might add a social sharing function (like in Piazza or other apps) that allows students to share their GenAI experience and skills with each other."} Students also proposed showing \textit{"examples of good vs. bad usage"} or \textit{"course-specific examples of when to use AI."}

\textbf{Technical Scaffolding.} Students envisioned AI tools that enforced learning rather than affording shortcuts. One proposed \textit{"If I were to re-design a Gen-AI tool for learning programming i would make it ask you questions before giving you the correct code, almost like a right of passage to make sure the person using it understands the concepts."} Another proposed screen time limits: \textit{"Maybe introduce a feature that will allow students to put like a screen limit on their GenAI."} These students recognized that the current process relied entirely on willpower without environmental design supporting intended behaviors, especially under time pressure.

\vspace{-1mm}

\section{Implications and Future Work}

Aligning with implementation intention research \cite{gollwitzer2006implementation}, the contract supported the majority of our students' agency in better GenAI decision-making, but for a subset of our students, metacognitive awareness did not sustain intentional practice against the tool's tendency towards productivity-based use. Even students who valued their guidelines abandoned them because GenAI's frictionless alternatives created a low-cost pull away from effortful learning. Sustaining intentionality across hundreds of daily programming decisions requires support beyond self-regulation. This suggests that supporting genuine student agency may sometimes mean helping students recognize that intentional use is not possible for them right now -- that the most agentic choice is deliberate non-use. 

For instructors considering this approach, the contract successfully scaffolds students without established GenAI practices, making implicit choices explicit. Sustaining that intentionality requires much willpower and reflection. Students' suggestions in \S \ref{subsection:suggestions} may be helpful in reducing this burden. The contract helped students recognize the mismatch between their learning goals and GenAI's affordances, but resisting the tool's defaults still required self-regulation capacities not all students had developed. As such, instructors should stress that non-use is still a valid outcome of the intervention, not a failure to integrate the tool. We adopted a harm reduction philosophy to support student agency rather than impose rules, but our findings suggest this may require work from both sides: educators providing scaffolding and tools designed to work \textit{with} students' goals rather than undermining them. The contract demonstrated that students can articulate these goals and reflect meaningfully on alignment. 
Our next step will test whether making these goals and reflections visible to the tools themselves can reduce the burden of self-regulation to sustainable levels.
If tools cannot adapt to the context of students' learning needs, their usefulness in programming education may remain limited to students who least need the support \cite{prather2024widening}.

\clearpage 

\bibliographystyle{ACM-Reference-Format}
\bibliography{base}

@String{Computing = "Computing" }

@String{Computer = "{IEEE} Computer" }

@String{Springer = "Springer-Verlag" }

@incollection{denny2024desirable,
  title={Desirable characteristics for ai teaching assistants in programming education},
  author={Denny, Paul and MacNeil, Stephen and Savelka, Jaromir and Porter, Leo and Luxton-Reilly, Andrew},
  booktitle={Proceedings of the 2024 on Innovation and Technology in Computer Science Education V. 1},
  pages={408--414},
  year={2024}
}

@article{gollwitzer2006implementation,
  title={Implementation intentions and goal achievement: A meta-analysis of effects and processes},
  author={Gollwitzer, Peter M and Sheeran, Paschal},
  journal={Advances in experimental social psychology},
  volume={38},
  pages={69--119},
  year={2006},
  publisher={Elsevier}
}

@article{yilmaz2023augmented,
  title={Augmented intelligence in programming learning: Examining student views on the use of ChatGPT for programming learning},
  author={Yilmaz, Ramazan and Yilmaz, Fatma Gizem Karaoglan},
  journal={Computers in Human Behavior: Artificial Humans},
  volume={1},
  number={2},
  pages={100005},
  year={2023},
  publisher={Elsevier}
}

@inproceedings{liu2024teaching,
  title={Teaching CS50 with AI: leveraging generative artificial intelligence in computer science education},
  author={Liu, Rongxin and Zenke, Carter and Liu, Charlie and Holmes, Andrew and Thornton, Patrick and Malan, David J},
  booktitle={Proceedings of the 55th ACM technical symposium on computer science education V. 1},
  pages={750--756},
  year={2024}
}

@inproceedings{liffiton2023codehelp,
  title={Codehelp: Using large language models with guardrails for scalable support in programming classes},
  author={Liffiton, Mark and Sheese, Brad E and Savelka, Jaromir and Denny, Paul},
  booktitle={Proceedings of the 23rd Koli Calling International Conference on Computing Education Research},
  pages={1--11},
  year={2023}
}

@inproceedings{kazemitabaar2024codeaid,
  title={Codeaid: Evaluating a classroom deployment of an llm-based programming assistant that balances student and educator needs},
  author={Kazemitabaar, Majeed and Ye, Runlong and Wang, Xiaoning and Henley, Austin Zachary and Denny, Paul and Craig, Michelle and Grossman, Tovi},
  booktitle={Proceedings of the 2024 chi conference on human factors in computing systems},
  pages={1--20},
  year={2024}
}

@inproceedings{zastudil2023generative,
  title={Generative ai in computing education: Perspectives of students and instructors},
  author={Zastudil, Cynthia and Rogalska, Magdalena and Kapp, Christine and Vaughn, Jennifer and MacNeil, Stephen},
  booktitle={2023 IEEE Frontiers in Education Conference (FIE)},
  pages={1--9},
  year={2023},
  organization={IEEE}
}

@inproceedings{hou2024effects,
  title={The effects of generative ai on computing students’ help-seeking preferences},
  author={Hou, Irene and Mettille, Sophia and Man, Owen and Li, Zhuo and Zastudil, Cynthia and MacNeil, Stephen},
  booktitle={Proceedings of the 26th australasian computing education conference},
  pages={39--48},
  year={2024}
}

@article{bandura1982self,
  title={Self-efficacy mechanism in human agency.},
  author={Bandura, Albert},
  journal={American psychologist},
  volume={37},
  number={2},
  pages={122},
  year={1982},
  publisher={American Psychological Association}
}

@inproceedings{parham2010empirical,
  title={Empirical evidence for the existence and uses of metacognition in computer science problem solving},
  author={Parham, Jennifer and Gugerty, Leo and Stevenson, DE},
  booktitle={Proceedings of the 41st ACM technical symposium on Computer science education},
  pages={416--420},
  year={2010}
}

@article{kohler2008abstinence,
  title={Abstinence-only and comprehensive sex education and the initiation of sexual activity and teen pregnancy},
  author={Kohler, Pamela K and Manhart, Lisa E and Lafferty, William E},
  journal={Journal of adolescent Health},
  volume={42},
  number={4},
  pages={344--351},
  year={2008},
  publisher={Elsevier}
}

@article{naisteter2010minimizing,
  title={Minimizing harm and maximizing pleasure: Considering the harm reduction paradigm for sexuality education},
  author={Naisteter, Michal A and Sitron, Justin A},
  journal={American Journal of Sexuality Education},
  volume={5},
  number={2},
  pages={101--115},
  year={2010},
  publisher={Taylor \& Francis}
}

@inproceedings{kirkpatrick2025challenging,
  title={Challenging AI as Critical Thinking},
  author={Kirkpatrick, Michael S},
  booktitle={International Conference on the Ethical and Social Impacts of ICT},
  pages={239--251},
  year={2025},
  organization={Springer}
}

@incollection{margulieux2024self,
  title={Self-regulation, self-efficacy, and fear of failure interactions with how novices use llms to solve programming problems},
  author={Margulieux, Lauren E and Prather, James and Reeves, Brent N and Becker, Brett A and Cetin Uzun, Gozde and Loksa, Dastyni and Leinonen, Juho and Denny, Paul},
  booktitle={Proceedings of the 2024 on Innovation and Technology in Computer Science Education V. 1},
  pages={276--282},
  year={2024}
}

@inproceedings{cunningham2022bringing,
  title={Bringing" High-Level" Down to Earth: Gaining Clarity in Conversational Programmer Learning Goals},
  author={Cunningham, Kathryn and Qiao, Yike and Feng, Alex and O'Rourke, Eleanor},
  booktitle={Proceedings of the 53rd ACM Technical Symposium on Computer Science Education-Volume 1},
  pages={551--557},
  year={2022}
}

@article{marlatt1996harm,
  title={Harm reduction: Come as you are},
  author={Marlatt, G Alan},
  journal={Addictive behaviors},
  volume={21},
  number={6},
  pages={779--788},
  year={1996},
  publisher={Elsevier}
}

@article{wang2025chatgpt,
  title={ChatGPT-enhanced self-regulated learning in programming education: impacts on motivation, self-efficacy, and learning outcomes},
  author={Wang, Zilin and Zou, Di and Zhang, Ruofei and Lee, Lap-Kei and Xie, Haoran and Wang, Fu Lee},
  journal={Interactive Learning Environments},
  pages={1--26},
  year={2025},
  publisher={Taylor \& Francis}
}

@inproceedings{ko2025student,
  title={Student perceptions of the help resource landscape},
  author={Ko, Shao-Heng and Stephens-Martinez, Kristin and Zahn, Matthew and Velasco, Yesenia and Battestilli, Lina and Heckman, Sarah},
  booktitle={Proceedings of the 56th ACM Technical Symposium on Computer Science Education V. 1},
  pages={596--602},
  year={2025}
}

@article{ko2025rethinking,
  title={Rethinking Computing Students’ Help Resource Utilization through Sequentiality},
  author={Ko, Shao-Heng and Stephens-Martinez, Kristin},
  journal={ACM Transactions on Computing Education},
  volume={25},
  number={1},
  pages={1--34},
  year={2025},
  publisher={ACM New York, NY}
}

@inproceedings{ali2025analysis,
  title={Analysis of generative AI policies in computing course syllabi},
  author={Ali, Areej and Collier, Aayushi Hingle and Dewan, Umama and McDonald, Nora and Johri, Aditya},
  booktitle={Proceedings of the 56th ACM Technical Symposium on Computer Science Education V. 1},
  pages={18--24},
  year={2025}
}

@article{padiyath2024have,
  title={Do I Have a Say in This, or Has ChatGPT Already Decided for Me?},
  author={Padiyath, Aadarsh},
  journal={XRDS: Crossroads, The ACM Magazine for Students},
  volume={31},
  number={1},
  pages={52--55},
  year={2024},
  publisher={ACM New York, NY, USA}
}

@incollection{prather2023robots,
  title={The robots are here: Navigating the generative ai revolution in computing education},
  author={Prather, James and Denny, Paul and Leinonen, Juho and Becker, Brett A and Albluwi, Ibrahim and Craig, Michelle and Keuning, Hieke and Kiesler, Natalie and Kohn, Tobias and Luxton-Reilly, Andrew and others},
  booktitle={Proceedings of the 2023 working group reports on innovation and technology in computer science education},
  pages={108--159},
  year={2023}
}

@incollection{vadaparty2024cs1,
  title={Cs1-llm: Integrating llms into cs1 instruction},
  author={Vadaparty, Annapurna and Zingaro, Daniel and Smith IV, David H and Padala, Mounika and Alvarado, Christine and Gorson Benario, Jamie and Porter, Leo},
  booktitle={Proceedings of the 2024 on Innovation and Technology in Computer Science Education v. 1},
  pages={297--303},
  year={2024}
}

@inproceedings{lau2023ban,
  title={From" Ban it till we understand it" to" Resistance is futile": How university programming instructors plan to adapt as more students use AI code generation and explanation tools such as ChatGPT and GitHub Copilot},
  author={Lau, Sam and Guo, Philip},
  booktitle={Proceedings of the 2023 ACM Conference on International Computing Education Research-Volume 1},
  pages={106--121},
  year={2023}
}

@inproceedings{prather2024widening,
  title={The widening gap: The benefits and harms of generative ai for novice programmers},
  author={Prather, James and Reeves, Brent N and Leinonen, Juho and MacNeil, Stephen and Randrianasolo, Arisoa S and Becker, Brett A and Kimmel, Bailey and Wright, Jared and Briggs, Ben},
  booktitle={Proceedings of the 2024 ACM Conference on International Computing Education Research-Volume 1},
  pages={469--486},
  year={2024}
}

@article{boguslawski2025programming,
  title={Programming education and learner motivation in the age of generative AI: student and educator perspectives},
  author={Boguslawski, Samuel and Deer, Rowan and Dawson, Mark G},
  journal={Information and Learning Sciences},
  volume={126},
  number={1/2},
  pages={91--109},
  year={2025},
  publisher={Emerald Publishing Limited}
}

@inproceedings{padiyath2024insights,
author = {Padiyath, Aadarsh and Hou, Xinying and Pang, Amy and Viramontes Vargas, Diego and Gu, Xingjian and Nelson-Fromm, Tamara and Wu, Zihan and Guzdial, Mark and Ericson, Barbara},
title = {Insights from Social Shaping Theory: The Appropriation of Large Language Models in an Undergraduate Programming Course},
year = {2024},
isbn = {9798400704758},
publisher = {Association for Computing Machinery},
address = {New York, NY, USA},
url = {https://doi.org/10.1145/3632620.3671098},
doi = {10.1145/3632620.3671098},
booktitle = {Proceedings of the 2024 ACM Conference on International Computing Education Research - Volume 1},
pages = {114–130},
numpages = {17},
location = {Melbourne, VIC, Australia},
series = {ICER '24}
}

@inproceedings{gandhi2025s,
  title={" That’s Not the Way I Would Explain It": A Teacher-Researcher’s Autoethnography of Learning to Program With ChatGPT},
  author={Gandhi, Abbey and Muldner, Kasia},
  booktitle={Proceedings of the 2025 ACM Conference on International Computing Education Research V. 1},
  pages={227--239},
  year={2025}
}

@article{zimmerman2002becoming,
  title={Becoming a self-regulated learner: An overview},
  author={Zimmerman, Barry J},
  journal={Theory into practice},
  volume={41},
  number={2},
  pages={64--70},
  year={2002},
  publisher={Taylor \& Francis}
}

@article{loksa2022metacognition,
  title={Metacognition and self-regulation in programming education: Theories and exemplars of use},
  author={Loksa, Dastyni and Margulieux, Lauren and Becker, Brett A and Craig, Michelle and Denny, Paul and Pettit, Raymond and Prather, James},
  journal={ACM Transactions on Computing Education (TOCE)},
  volume={22},
  number={4},
  pages={1--31},
  year={2022},
  publisher={ACM New York, NY}
}

@article{gollwitzer1999implementation,
  title={Implementation intentions: strong effects of simple plans.},
  author={Gollwitzer, Peter M},
  journal={American psychologist},
  volume={54},
  number={7},
  pages={493},
  year={1999},
  publisher={American Psychological Association}
}

@inproceedings{bernstein2025beyond,
  title={Beyond the Benefits: A Systematic Review of the Harms and Consequences of Generative AI in Computing Education},
  author={Bernstein, Seth and Rahman, Ashfin and Sharifi, Nadia and Terbish, Ariunjargal and MacNeil, Stephen},
  booktitle={Proceedings of the 25th Koli Calling International Conference on Computing Education Research},
  pages={1--18},
  year={2025}
}


\end{document}